\documentclass[fleqn,usenatbib]{mnras}

\usepackage{newtxtext,newtxmath}
\usepackage[T1]{fontenc}
\usepackage{bm}

\DeclareRobustCommand{\VAN}[3]{#2}
\let\VANthebibliography\thebibliography
\def\thebibliography{\DeclareRobustCommand{\VAN}[3]{##3}\VANthebibliography}

\usepackage{graphicx}	% Including figure files
\usepackage{amsmath}	% Advanced maths commands
\usepackage{acronym}

\acrodef{NS}{Neutron Star}
\acrodef{BH}[BH]{Black Hole}
\acrodef{BBH}[BBH]{Binary Black Hole}
\acrodef{GW}[GW]{gravitational wave}
\acrodef{PE}{Parameter Estimation}
\acrodef{MC}{Monte Carlo}
\acrodef{MCMC}{Markov Chain Monte Carlo}
\acrodef{DOF}{degrees of freedom}
\acrodef{PSD}{power spectral density}
\acrodef{SNR}{signal-to-noise ratio}

\newcommand{\D}{\mathrm{d}}

\title[]{What's in a binary black hole's mass parameter?}

\author[V.~Tiwari]{
Vaibhav Tiwari$^{1}$\thanks{E-mail: tiwariv@cardiff.ac.uk}
\\
$^{1}$Gravity Exploration Institute, School of Physics and Astronomy, Cardiff University, Queens Buildings, The Parade Cardiff CF24 3AA, UK.
}
\pubyear{2023}

\begin{document}
\label{firstpage}
\pagerange{\pageref{firstpage}--\pageref{lastpage}}
\maketitle

\begin{abstract}
The \ac{BH} masses measured from gravitational wave observations appear to cluster around specific mass values. Consequently, the primary~(and chirp) mass distribution of \acp{BBH} inferred using these measurements shows four emerging peaks. These peaks are approximately located at a primary~(chirp) mass value of 10$M_\odot$~(8$M_\odot$), 20$M_\odot$~(14$M_\odot$), 35$M_\odot$~(28$M_\odot$) and 63$M_\odot$~(49$M_\odot$). Although the presence of the first and third peaks has been attributed to \ac{BBH} formation in star clusters or due to the evolution of stellar binaries in isolation, the second peak has received relatively less attention because it lacks significance in the primary mass distribution. In this article, we report that confidence in the second peak depends on the mass parameter we choose to model the population on. Unlike primary mass, this peak is significant when modelled on the chirp mass. We discuss the disparity as a consequence of mass asymmetry in the observations that cluster at the second peak. Finally, we report this asymmetry as part of a potential trend in the mass ratio distribution manifested as a function of the chirp mass, but not as a function of primary mass, when we include the observation GW190814 in our modelling. The chirp mass is not a parameter of astrophysical relevance. Features present in the chirp mass, but not in the primary mass, are relatively difficult to explain and expected to garner significant interest.
\end{abstract}


%%%%%%%%%%%%%%%%%%%%%%%%%%%%%%%%%%%%%%%%%%%%%%%%%%

%%%%%%%%%%%%%%%%% BODY OF PAPER %%%%%%%%%%%%%%%%%%

\section{Introduction}

Gravitational wave observations are beginning to inform on the population of the \acp{BBH}. The most recent report by the LIGO-Virgo-Kagra (LVK) collaboration inferred the \ac{BBH} population~\citep{o3b_rnp} using the 69 \ac{BBH} observations from their third \ac{GW} transient catalog~\citep{o3b_cat}. Although still in the initial stages and limited by the number of observations, the \ac{BBH} population has presented a complex picture in terms of the mass and spin distributions~\citep{2022ApJ...928..155T}. The mass ratio and spins distributions vary with the mass distribution and remain a challenge to be consistently explained by the proposed formation channels\footnote{Chirp mass $\mathcal{M} = (m_1+m_2)^{3/5} / (m_1+m_2)^{1/5}$ dominates the phase evolution of GWs. Mass-ratio is the ratio of secondary mass~($m_2$) and the primary mass~($m_1$), defined as $q=m_2/m_1$. Aligned spins are the components of the spins aligned with the orbital angular momentum.}. Among others, efforts have focused on independently investigating the spin~\citep{2022ApJ...928...75H, 2022arXiv221212113B}, the mass-ratio~\citep{2022ApJ...933L..14L, 2022PhRvD.106b3014S}, correlations in the mass-spin~\citep{2021ApJ...922L...5C, 2022PhRvD.105l3024F, 2022MNRAS.517.3928A, 2022ApJ...941L..39W, 2022MNRAS.517.2738M}, mass-redshift~\citep{2022ApJ...935..126B} and the spin-redshift distributions~\citep{2022ApJ...932L..19B, 2022A&A...665A..59B}. 

The \ac{GW} observation GW190814 adds to the challenges as it does not fit with the rest of the \ac{BBH} population~\citep{2019ApJ...882L..24A, 2020ApJ...896L..44A}. GW190814's  primary object with a mass of around 23$M_\odot$ is expected to be a \ac{BH}, however, the secondary object with a mass of 2.6$M_\odot$ lies in the lower mass gap\footnote{A predicted absence of compact objects with masses in between the most massive neutron stars and least massive black holes with masses $\sim$2--5$M_\odot$.}. The secondary object has been interpreted as either the heaviest \ac{NS}~\citep{2021PhRvC.103b5808D, 2022PhRvD.105k4042I} or the lightest \ac{BH}~\citep{2020MNRAS.499L..82M, 2020PhRvC.102f5805F, 2021ApJ...908L...1T} ever discovered in a double-compact system. Its observation has been explained or investigated in many ways, such as, it being a \ac{BH} of primordial origin~\citep{2020arXiv200706481C, 2021PhRvL.126e1302J, 2022PhRvD.106l3526F}. Future observations will help reveal the origin of GW190814, but there lacks a concrete proposal at the current moment. We note that the chirp mass of GW190814 lies in the range $4M_\odot - 10M_\odot$~\citep{o3b_cat}, a point we will revisit later in this article.

In addition, \ac{BH} masses and spins inferred from x-ray binary measurements and \ac{GW} measurements are possibly in tension~\citep{2022ApJ...929L..26F}. The tension in mass has been suggested to arise because the two methods probe environments that differ in metallicity that are expected to produce \acp{BH} differing in masses~\citep{2021arXiv211109401B, 2022arXiv221001825L}, and the tension in spins should be assessed while keeping the uncertainties and systematics the \ac{BH} spin measurements from x-ray binaries are subjected to~\citep{2021arXiv211109401B} or the possibility of their formation through alternate scenarios~\citep{2022ApJ...938L..19G}.

In contrast to these challenges, the \ac{BBH} mass distribution shows an organised structure. There is a presence of several peaks, suggesting the over-production of binaries with masses clustered around specific values. Earlier, we reported the locations of the peaks seem to bear a constant factor~\citep{2021ApJ...913L..19T, 2022ApJ...928..155T}. The first peak is located at an approximate chirp mass value of 8$M_\odot$~(10$M_\odot$ in primary mass)~\citep{2021ApJ...913L..19T, 2021ApJ...922..258V, 2022ApJ...924..101E, 2022PhRvD.105l3014S} and has been predicted to originate from the \acp{BBH} mergers formed from the evolution of stellar binaries in isolation~\citep{o3b_rnp, 2022ApJ...940..184V, 2023arXiv230502380S}. The second peak is located at an approximate chirp mass value of 14$M_\odot$~(20$M_\odot$ in primary mass)~\citep{2021ApJ...913L..19T, o3b_rnp, 2022ApJ...928..155T, 2022arXiv221012834E}. The second peak has been suggested to be only marginally~\citep{2022arXiv220712409W, 2023arXiv230207289C, 2023arXiv230100834F} or moderately significant~\citep{o3b_rnp} and recent works suggest it due to merger of \ac{BBH} from isolated binaries~\citep{2023arXiv230502380S}. The third peak is located at an approximate chirp mass value of 28$M_\odot$~(35$M_\odot$ in primary mass)~\citep{2019ApJ...882L..24A} and the suggested mechanism for its formation includes pile-up of \ac{BBH} near the cut-off mass for the pair-instability supernovae~\citep{2018ApJ...856..173T}, and dynamical formation in globular clusters~\citep{2022arXiv220801081A}. The fourth peak is located at an approximate chirp mass value of 49$M_\odot$~(63$M_\odot$ in primary mass). This peak is only marginally significant. Multiple peaks have also been proposed due to repeated mergers of \acp{BH} in star clusters or active galactic nuclei~\citep{2022arXiv220905766M, 2023PhRvD.107f3007L}.

The astrophysical processes responsible for the formation of \ac{BBH} are anticipated to imprint the mass spectrum with features. Current proposals suggest the formation of compact binaries from the evolution of massive binary stars~(e.g. see ~\citet{2022Galax..10...76S}) and in dense star clusters due to many-body interaction~(e.g. see~\citet{2016PhRvD..93h4029R, 2021MNRAS.505..339M, 10.1093/mnras/stac1163, 2023arXiv230211613F}), including binary formation in active galactic nuclei~(e.g. see~\cite{2023arXiv230104187G, 2023arXiv230214071A}). The first and the third peaks in the mass distribution have been jointly focused on by recent works~(e.g. see~\citet{2022ApJ...941L..39W, 2023arXiv230401288G}). However, the second peak, because it is of marginal presence in primary mass, has not been confidently explained or received attention.

In this article, we report the significance of the peaks in the mass distribution.  In particular, we investigate how likely is it for the peaks to arise from a featureless mass distribution. We also show that the significance of the second peak depends on the mass parameter used in modelling the population. When modelling the population on the chirp mass, we conclude the second peak to be significant, but when modelling the population on the primary mass, we conclude a marginal significance. We show that the mass parameters give different confidence in the peaks because of mass asymmetry at the second peak. Finally, we report this asymmetry may be a part of a potential trend in the mass ratio that is manifested when we include the observation GW190814 in the analysis. We summarise the methodology and report the results in Sec.~\ref{sec:results} and discuss a trend in mass asymmetry in Sec.~\ref{sec:trend}.
\section{Summary of the Method and results}
\label{sec:results}
A mass parameter in a two-body problem can be chosen in different ways. Some of the examples relevant to \acp{BBH} include, i) component mass: any of the two masses without making a distinction, ii) the primary mass: heavier of the two \acp{BH}, iii) total mass: the sum of the two masses, and iv) chirp mass: the most accurately measured function of the two masses. Along with the chosen parameter, one needs also to specify the ratio between the two masses to uniquely determine the masses of both components. Due to their astrophysical relevance primary mass and mass ratio are often the parameter of choice when inferring the \ac{BBH} population. However, once inferred the distribution on any other mass parameter can be obtained by performing a parameter transformation. It requires a simple application of a relevant Jacobian matrix. If the population is composed of comparable mass \acp{BBH}, the  resulting Jacobian will be mostly independent of the mass ratio~\citep{2021arXiv210409508C}. The one-dimensional distribution of the new mass parameter will only be a scaled version of the one-dimensional distribution of the old mass parameter, i.e. the distribution of the new parameter will be similar to the old one but with the x/y axis redefined. Any feature present in the old mass parameter will also appear at a comparable confidence in the new mass parameter. But, if the mass ratio has a dependence on the mass parameter, the Jacobian will also show a dependence on the mass parameter resulting in different scaling at different mass values. Any feature present in the old mass parameter may appear at a lower or higher confidence in the new mass parameter. 

The mass ratio distribution of \ac{BBH} has shown mass dependence~\citep{2022ApJ...928..155T}. In this article, we focus on an emerging structure in the \ac{BBH} mass distribution and investigate if the confidence in features depends on the chosen mass parameter. We use the mixture model framework, Vamana, that can infer all the major one- and two-dimensional features in the \ac{BBH} population~\citep{2021CQGra..38o5007T}. In practice we can infer the population using one mass parameter and draw conclusions by making parameter transformations. Instead, we infer the \ac{BBH} population using several modifications of Vamana that either directly model the primary mass or the chirp mass. Primary mass is a parameter of astrophysical importance and its population-level distribution is expected to be directly impacted by the physics of binary formation and merger. On the other hand, chirp mass does not have direct astrophysical relevance. It is an interesting parameter because it dominates the phase evolution of a binary and consequently is the most accurately measured mass parameter~\citep{1993PhRvD..47.2198F, Cutler:1994ys}. However, any significant feature observed in the chirp mass distribution can not be ignored. Either such a feature needs to be explained as arising from a unique combination of other astrophysically motivated parameters or justified as an outcome of the underlying systematics of the model used in inferring the distribution. A brief description of these population models is as follows, a detailed description is presented in Appendix~\ref{method}. All models use a mixture of weighted components to infer the population. The components contain
\begin{enumerate}
    \item Model $\mathbb{M}_{\mathcal{M}}$: a truncated bi-variate normal to model the chirp mass and mass ratio distribution.
    \item Model $\mathbb{M}_{m_1}$: a truncated bi-variate normal to model the primary mass and mass ratio distribution.
    \item Model $\mathbb{M}^{pl}_{m_1}$: a uni-variate normal to model the primary mass distribution. A single power law is used to model the mass ratio of the full mass range.
\end{enumerate}
The bi-variates are truncated because the mass ratio can obtain a maximum value of one. In addition, for all the models, the components also include a uni-variate normal to identically but independently model the aligned spin distribution, and a power law to model the redshift evolution of the merger rate.
The first two population models are Gaussian mixtures and are expected to model a range of distributions in masses and spins. The third model, $\mathbb{M}^{pl}_{m_1}$, uses a single power law to infer the mass ratio for the full population while flexibly modelling the primary mass distribution. This population model is comparative to other approaches that flexibly model the primary mass and use a phenomenological model to infer the mass ratio~\citep{o3b_rnp, 2022arXiv221012834E}. 

\begin{center}
\begin{figure*}
\includegraphics[width=0.9\textwidth]{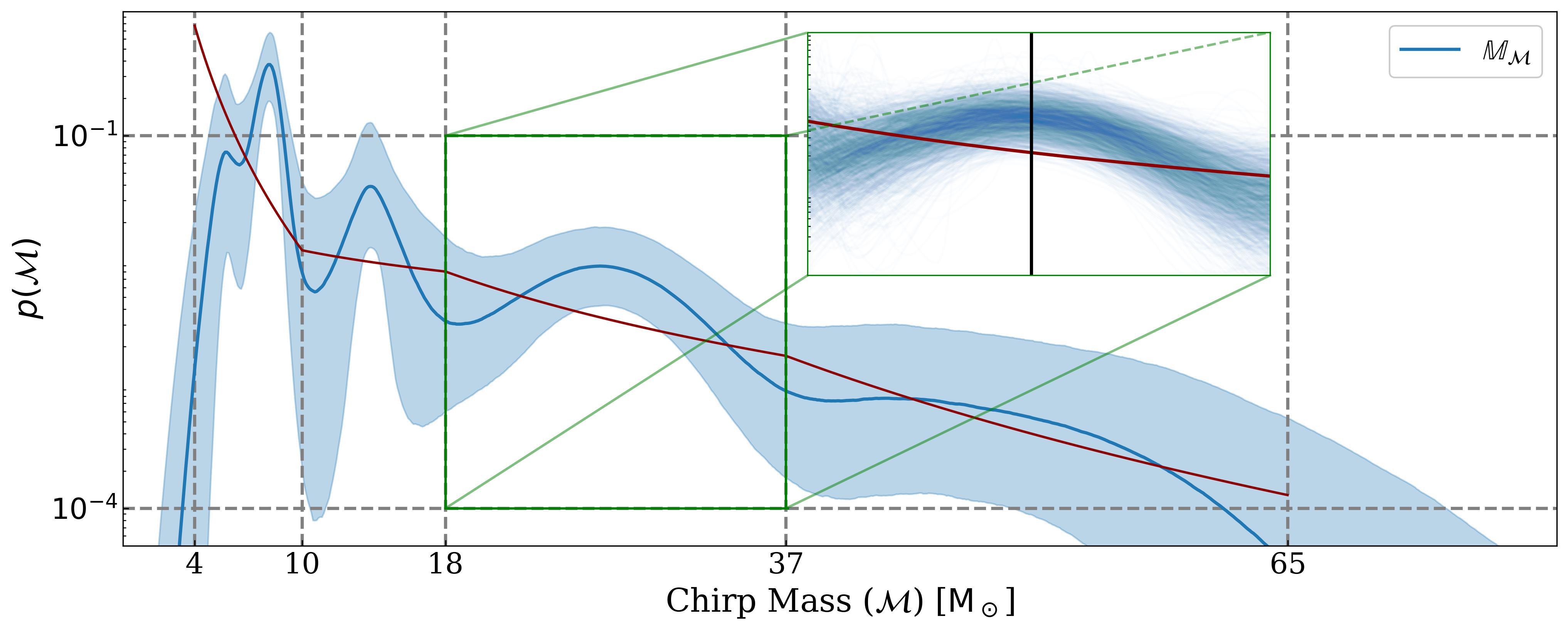}
\includegraphics[width=0.9\textwidth]{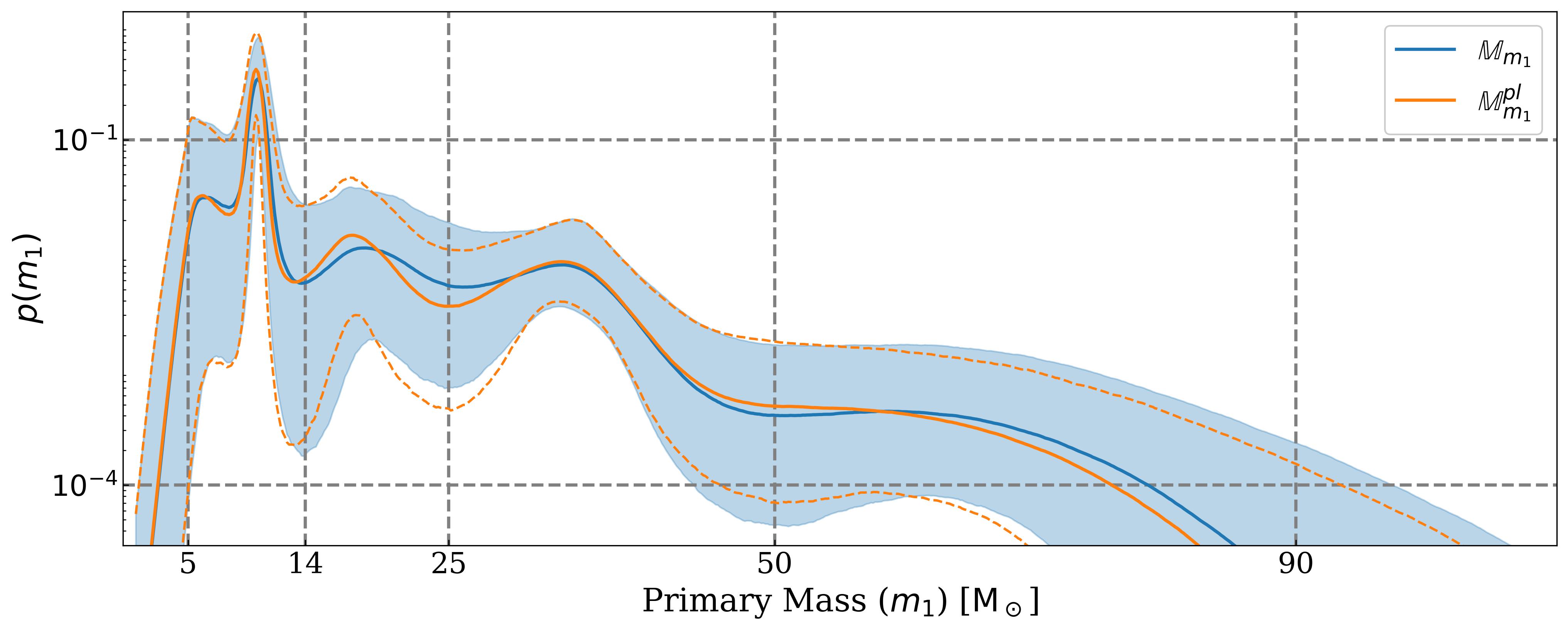}
\vspace{-1em}
\caption{Top) chirp mass distribution inferred using model $\mathbb{M}_{\mathcal{M}}$. The solid blue line is the density and the blue band is the 90\% credible interval. The vertical grid lines identify the four chirp mass segments. The dark red line is a power law distribution that minimises over or under-density throughout the mass range. The inset focuses on the third peak and shows different draws of the inferred chirp mass distribution~(see Eq.~\ref{eq:post_pred}). The black line indicates the location of the highest over-density, bottom) primary mass distribution inferred using models $\mathbb{M}_{m_1}$ and $\mathbb{M}^{pl}_{m_1}$. The solid blue line is the median density for model $\mathbb{M}_{m_1}$ and the bands are the 90\% credible interval. The solid orange line is the median density for model $\mathbb{M}^{pl}_{m_1}$. To improve visual presentation we plot the 90\% credible intervals using dashed orange lines. The dashed red lines identify the four primary mass segments.}
\label{fig:diffmerger}
\end{figure*}
\end{center}

Fig.~\ref{fig:diffmerger} plots the inferred distribution over the mass range. The median of the inferred mass distribution shows the presence of four peaks for all three models. However, the median should be viewed while considering the 90\% credible interval as shown by the blue band in Fig.~\ref{fig:diffmerger}. The blue band depicts the range of possibilities the distribution can acquire. A larger interval can easily accommodate a featureless distribution. Thus, the fourth peak which has a large credible interval is not expected to be significant. Visually, the prominence of the second peak decreases in orders $\mathbb{M}_{\mathcal{M}}$, $\mathbb{M}^{pl}_{m_1}$, and $\mathbb{M}_{m_1}$ for our three models. The inferred primary mass distribution using $\mathbb{M}_{m_1}$ and $\mathbb{M}^{pl}_{m_1}$ is mostly consistent. The difference between them occurs around the second peak. However, the model $\mathbb{M}^{pl}_{m_1}$ is disfavoured against the model $\mathbb{M}_{m_1}$ by a Bayes factor of 10.\footnote{Vamana uses the Metropolis-Hastings~\citep{10.1093/biomet/57.1.97} algorithm to sample the hyper-parameter posterior but it can estimate an approximate value of the marginal likelihood/evidence.} The mass ratio distribution varies with the masses of the binaries~\citep{2022ApJ...928..155T}, thus modelling the full mass range using one power law does not fit the data well for the model $\mathbb{M}^{pl}_{m_1}$. On the other hand, $\mathbb{M}_{m_1}$ can model flexibly throughout the mass range as each component models the mass ratio independently. 

Next, we estimate the confidence in the peaks. We estimate this confidence by asking the question: how likely is it for the inferred peaks to originate from  a distribution that is featureless? As the distribution of a handful of observations on the mass range follows a Poisson process, Gaussian mixtures will often infer local maxima at random mass values even if the distribution being inferred is featureless. One way to assess a model's tendency to infer a local maxima is by compiling statistics using simulated data~\citep{2022PhRvD.105l3014S, 2023arXiv230100834F}. For the presented analyses we follow the following methodology:
\begin{enumerate}
    \item Underlying distribution: approximate the astrophysical mass distribution using a distribution that does not contain a local maxima. We use a monotonically decaying broken power-law as the approximating distribution and our conclusions are based on this choice. Inference from our three population models yield three featureless mass distribution as described later. Moreover, our approximating distribution ignores mass ratio and spins by fixing them to a fiducial distribution as described in Appendix~\ref{sec:sims}.
    \item Synthetic Observations: From the broken power-law generate synthetic observations. The number of synthetic observations is the same as the number of observed \acp{BBH}; assign measurement uncertainty as described in Appendix~\ref{sec:sims}. By fixing the mass-ratio and spin distribution our synthetic population does not have a correlation between the parameters and does not reflect favorably or unfavourably on how synthetic observations cluster in the mass range. A full analysis with all the parameters included is possible but we expect it to be computationally expensive.
    \item Infer mass distribution: perform 1,000 simulations and record how often the inferred distribution is over-dense compared to the underlying broken power-law. Our simulated data is one-dimensional, thus only the location and scales of Gaussians modelling the mass distribution contribute to the likelihood~(distributions of mass ratio and spins are already fixed).
    \item Statistics: compile statistics that quantify over-densities inferred by Gaussian mixtures from the simulated data. 
    \item Confidence in peaks: finally, calculate the over-density of the peaks in the \ac{BBH} mass distribution. Estimate confidence in them by recording the fraction of simulations that produced a comparable over-density from the featureless distribution.
\end{enumerate}

 We quantify an over- or under-density at a given value of the mass parameter, $m$, by calculating the fraction, 
\begin{align}
f(m) =& \frac{\textrm{Number of posterior samples inferring an over-density}}{\textrm{Total number of posterior samples}} \nonumber \\
=& \frac{1}{n}\sum_{i=1}^{n}\; \chi\left(p_i(m) > p_\mathrm{underlying} (m)\right)
\label{eq:f}
\end{align}
where $p_i(m)$ is one of the several draws for the inferred mass distribution, $\chi$ is one when the enclosed condition is met for the $i^{\mathrm{th}}$ posterior samples and zero otherwise, and $n$ is the number of hyper-parameter posteriors sampled by the analysis.
A value of 0.5 indicates neither an over nor under-density. The larger or smaller the value of $f$, the larger the inferred population's departure from the underlying distribution and thus the larger the over or the under-density. 
\begin{center}
\begin{figure}
\includegraphics[width=0.49\textwidth]{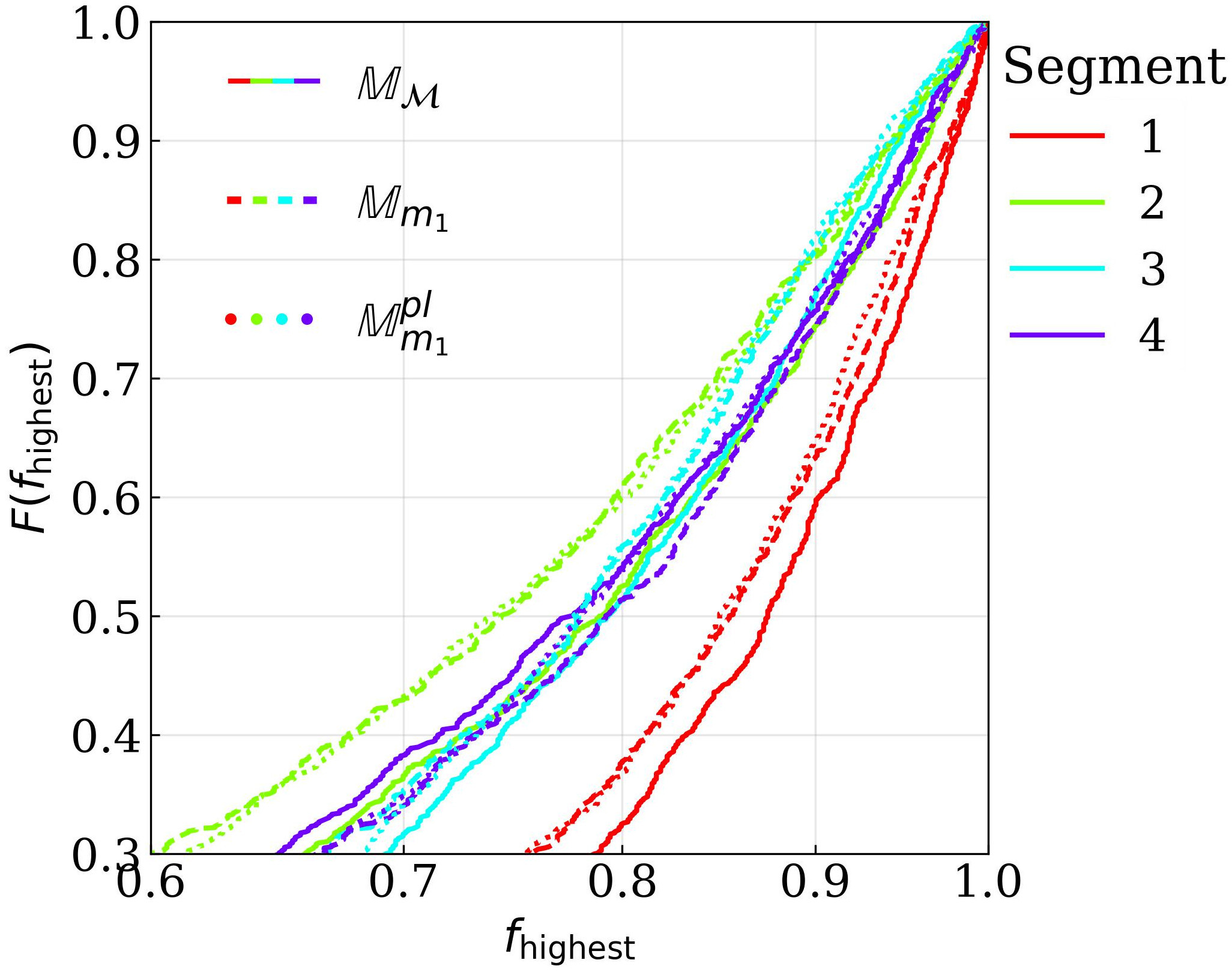}
\caption{Distribution of $f_\mathrm{highest}$, a measure of highest over-density for the featureless distributions. The featureless distributions do not contain a local maxima and best describe the mass distribution inferred by the three population models; thus three featureless distributions for the three population inference. We perform 1,000 simulation runs as described in Appendix~\ref{sec:sims}. Each simulation generates different realisations of data from a featureless distribution and infers the distribution by using the standard methodology. For each simulation,
we use Eq.~\ref{eq:f} to estimate $f_\mathrm{highest}$ for the four mass segments. The y-axis plots the fraction of simulations with over-density equal to or greater than $f_\mathrm{highest}$ corresponding to the value on the x-axis. The best-fit power-law has a steep decay in the first segment~(please see Appendix~\ref{sec:sims}) and combined with a sudden truncation at lower masses this segment results in the largest values of $f_\mathrm{highest}$. The x-axis has been terminated below 0.6 for the sake of clarity.}
\label{fig:f}
\end{figure}
\end{center}

For the peaks observed in the \ac{BBH} mass distribution, the underlying distribution is not known. As we are focusing on the local maxima, we split the mass range into segments~(4.0$M_\odot$, 10.0$M_\odot$, 18$M_\odot$, 37$M_\odot$, 65$M_\odot$), and for the primary distribution, we choose (5.0$M_\odot$, 14$M_\odot$, 25$M_\odot$, 50$M_\odot$, 90$M_\odot$). These values are the locations of local minima in the median of the inferred mass distribution. To assess how conveniently can we accommodate a featureless distribution in the confidence band, we place a monotonically decreasing broken power-law that connects the endpoint values. Using this broken power-law as the underlying distribution we calculate $f$ and vary the four exponents iteratively until we minimises $\int (0.5 - f(m))^2\;\mathrm{d}m$. The best-fit broken power-law departs least from $f=0.5$ throughout the mass range. The red curve shown in Fig.~\ref{fig:diffmerger} is this broken power-law for the chirp mass distribution. The inset shows different possibilities of the inferred chirp mass distribution inside the third segment. The black line indicates $f_{\mathrm{highest}}$, the mass value with the highest over-density for the best-fit power law. The value of $f_{\mathrm{highest}}$ in a segment quantifies the largest departure from the underlying distribution and consequently the presence of a peak. We summarise the values and locations of the peaks in Table~\ref{tab:peaks_summary} for the three models. To estimate the confidence in the peaks, we reverse the analysis and ask, how often $f_{\mathrm{highest}}$ measured in the segments can be generated by the broken power-law. Thus we use this broken power-law as the underlying featureless distribution and perform simulations. For each simulation, we record $f_{\mathrm{highest}}$ for the four segments. Fig.~\ref{fig:f} plots the distribution of $f_{\mathrm{highest}}$ for the three models.

\def\arraystretch{1.2}
\begin{table}
	\centering
	\caption{Location of highest over-density, $f_{\mathrm{highest}}$, for the three models and each mass segment. Model $\mathbb{M}_{\mathcal{M}}$ infers the chirp mass, and models $\mathbb{M}_{m_1}$ and $\mathbb{M}^{pl}_{m_1}$ infer the primary mass distribution.}
	\begin{tabular}{lcccc} % four columns, alignment for each

  \hline\
      Peak attributes & (i) & (ii) & (iii) & (iv) \\
      \hline
      Location ($\mathbb{M}_{\mathcal{M}}$) [$M_\odot$] & 8.3 & 13.9 & 28.1 & 49.0\\
      $f_{\mathrm{highest}}$ & 1 & 0.98 & 0.99 & 0.66\\
      \% sims with bigger $f_{\mathrm{highest}}$ & 0 & 6 & 2 & 73\\
      &&&&\\
      Location ($\mathbb{M}_{m_1}$) [$M_\odot$] & 10.3 & 20.0 & 34.5 & 64.4 \\
      $f_{\mathrm{highest}}$ & 0.97 & 0.68 & 0.98 & 0.32\\
      \% sims with bigger $f_{\mathrm{highest}}$ & 3 & 60 & 2 & 68\\
      &&&&\\
      Location ($\mathbb{M}^{pl}_{m_1}$) [$M_\odot$] & 10.2 & 18.3 & 34.8 & 62.9 \\
      $f_{\mathrm{highest}}$ &  0.99 & 0.82 &  0.99 & 0.68\\
      \% sims with bigger $f_{\mathrm{highest}}$ & 1 & 27 & 1 & 70\\
  \hline
	\end{tabular}
 \label{tab:peaks_summary}
\end{table}
\def\arraystretch{1.}
The confidence in the peaks can be estimated by comparing the $f_\mathrm{highest}$ values listed in Table~\ref{tab:peaks_summary} with the distribution we obtain from the simulations shown in Fig.~\ref{fig:f}. The shown distribution depends on various simplistic assumptions made in the featureless model~(described in Sec.~\ref{sec:sims}). However, the most important aspect is the bandwidth of the modelling functions. In the case of Gaussian mixtures, it is directly related to the maximum scale the Gaussians can acquire. For example, when modelled using Gaussians using narrow scales, the population model will create significant local maxima around all the observations. Thus, we have used the same methodology when inferring \ac{BBH} mass distribution from the various simulations  generated from the featurelss mass distribution. On making small changes to the featureless distribution we observe the distribution of $f_\mathrm{highest}$ remains largely unchanged. Overall, we can derive the following important takeaways:
\begin{itemize}
    \item Inferred by the model $\mathbb{M}_{\mathcal{M}}$, the first three peaks are significant in the chirp mass distribution. Less than 6\% of simulations were inferred with an over-density of $f_{\mathrm{highest}} > 0.98$.
    \item Inferred by model $\mathbb{M}_{m_1}$, which uses Gaussian mixture to infer the mass ratio distribution, the first and third peaks are significant. The significance of the second peak is marginal. More than 60\% simulations were inferred with an over-density of $f_{\mathrm{highest}} > 0.68$. A similar conclusion was drawn in \cite{2022arXiv220712409W}, \cite{2023arXiv230100834F} and \cite{2023arXiv230207289C}.
    \item Inferred by model $\mathbb{M}^{pl}_{m_1}$, which uses a single power law to infer the mass ratio of the full population,  the first and third peaks are significant. The significance of the second peak is bigger than inferred by the model $\mathbb{M}_{m_1}$. Around 30\% of simulations were inferred with an over-density of $f_{\mathrm{highest}} > 0.82$. A comparable conclusion was drawn in \cite{o3b_rnp}.
    \item The significance for all the peaks, except the second peak, remains almost the same for all three models. Models provide disparate conclusions on the significance of the second peak.
\end{itemize}
The featureless distribution independently focus on either the chirp mass or the primary mass distributions. However, the conclusions listed above are robust against parameter transformations, which means, the peaks are not an artefact of change in mass parameters. We have verified, that when using the model $\mathbb{M}_{\mathcal{M}}$ in inferring the chirp mass distribution from simulated data generated from featureless distribution that best represents the observed primary mass and an arbitrary mass ratio distribution~(for our test we drew mass ratio inferred for the second peak) the distribution of $f_\mathrm{highest}$ is very similar to the left plot of Fig.~\ref{fig:f}. 

\begin{center}
\begin{figure*}
\includegraphics[width=0.47\textwidth]{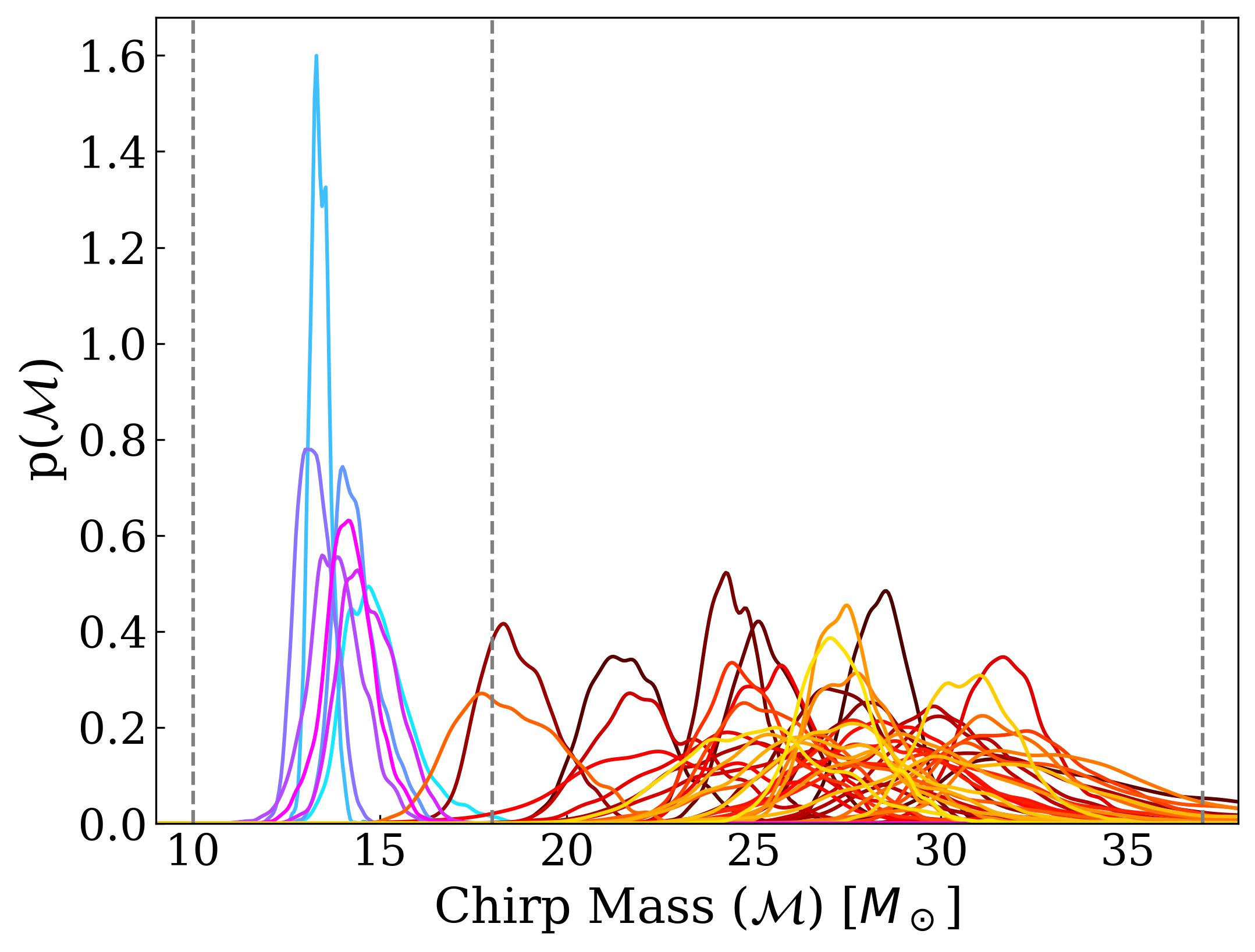}
\includegraphics[width=0.48\textwidth]{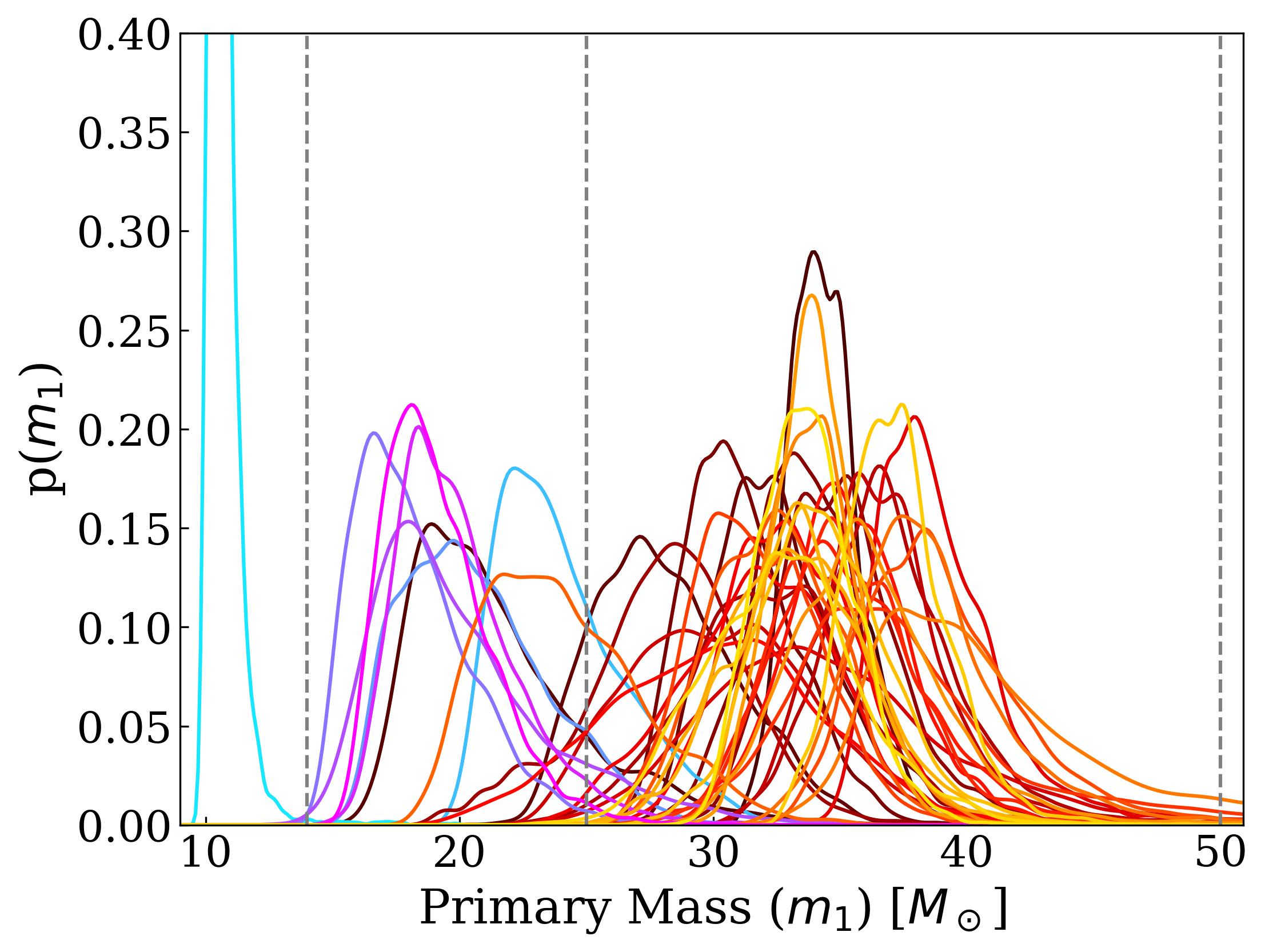}
\vspace{-1em}
\caption[Caption for LOF]{The chirp mass and primary mass estimates of observations that contribute to the second and the third peak. The dashed lines identify the second and third mass segments that enclose the peaks. Cool/Hot colours are used to plot estimates of individual observations that contribute to the second/third peak. For the sake of assigning colours, we allocate an observation to the second/third peak if the mean mass value lies inside the second/third mass segment. The two sets of clusters can be well identified in the chirp mass distribution. The second peak is less distinguishable in the primary mass estimates~(one observation has a long tail such that the mean primary mass lies in the second mass segment but the peak of the distribution is outside the segment.)}
\label{fig:pe_peaks}
\end{figure*}
\end{center}

These conclusions become apparent on reviewing the chirp and primary mass estimates of the observations that contribute to the second and third peaks. Fig.~\ref{fig:pe_peaks} shows estimated chirp and primary mass distributions for the observations that contribute to the second and third peaks. Two separated clusters are visible in the chirp mass. For primary mass, the observations at the second peak show a skew in distributions. This skew is primarily because overall these observations have a larger mass asymmetry than the observations contributing to the third peak. Arguing from the point of view of a significant peak present in the chirp mass distribution but not in the primary mass distribution, the mass asymmetry foundationally arises because of a correlation present between the primary mass and the mass ratio. On making draws from the estimated primary mass and mass ratio for each of the seven observations that contribute to the second peak in chirp mass, we obtain a Pearson correlation coefficient between the two parameters in the range of 0.82 and 0.98 at 90\% confidence. On the contrary, any positive correlation between two uncorrelated parameters is expected to be between 0 and 0.67. Such a large correlation requires a variation in mass ratios and thus results in mass asymmetry at the second peak.

\section{A trend in the mass ratio distribution?}
\label{sec:trend}

An astrophysical process may create \acp{BH} in a narrow primary or component mass range. For example, most of the \ac{BH} masses inferred from x-ray binaries lie in the range 5$M_\odot$--12$M_\odot$ and follow a distribution that peaks around $7M_\odot$~\citep{2016A&A...587A..61C}.  It is an interesting finding that the significance of the second peak depends on the mass parameter. A bigger significance when modelling on the chirp mass and a marginal significance when modelling on the primary mass requires a correlation between primary mass and mass ratio.

\begin{center}
\begin{figure}
\includegraphics[width=0.47\textwidth]{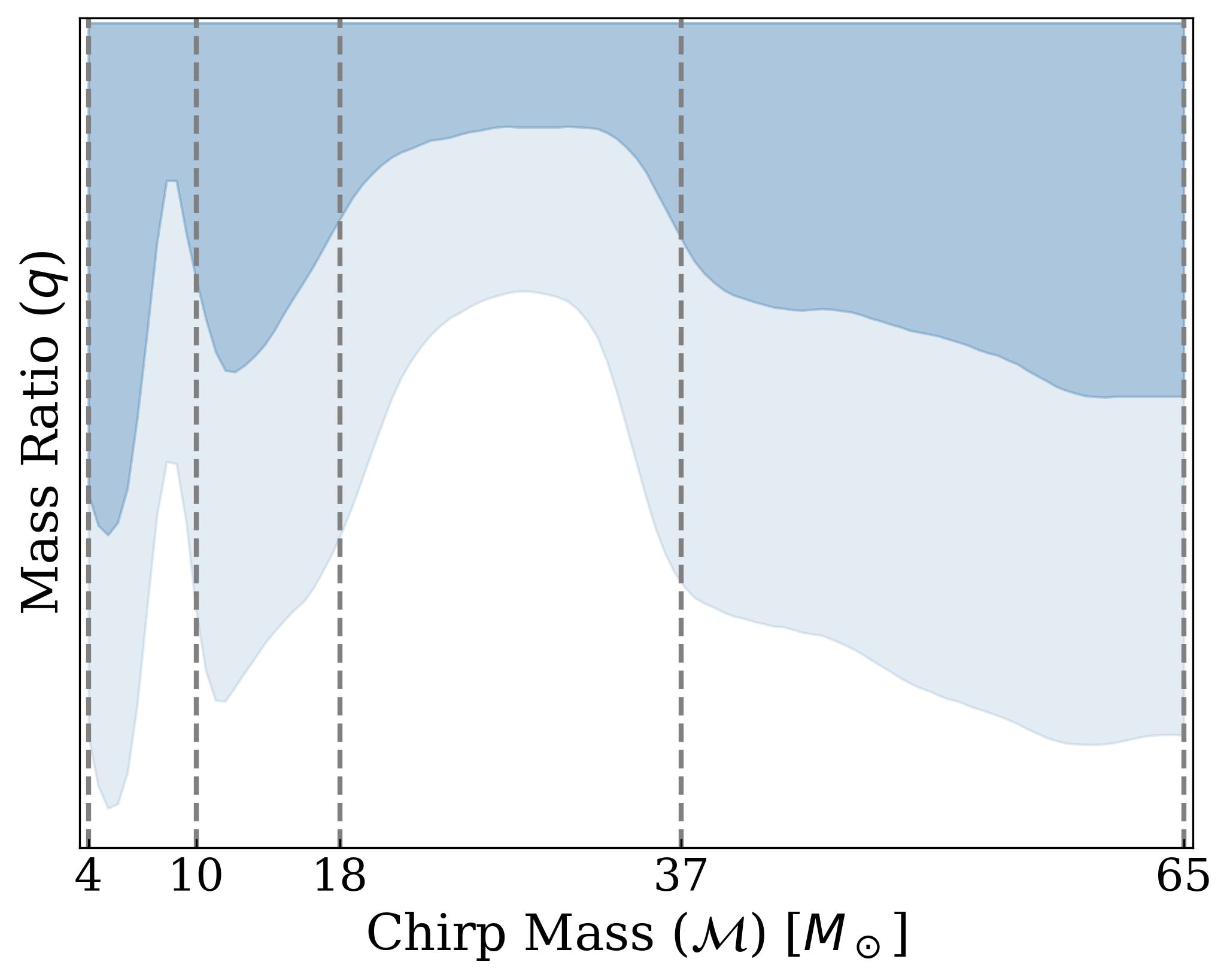}
\caption{Variation of mass ratio as a function of chirp mass. The dark/light bands show $50^\mathrm{th}/10^\mathrm{th}$ percentile. The first three peaks show increased mass asymmetry for smaller values of the chirp mass. The vertical grid lines show the segments that enclose the peaks. The asymmetry in the first segment is largely due to the observations GW190814 and GW200210\_092254. Their average chirp mass value is approximately 6.3$M_\odot$. The largest asymmetry occurs at a slightly smaller value. This is an artefact introduced when correcting for the selection effect using a limited number of injections~(see Eq.~\ref{eq:pop_bayes})~\citep{2022arXiv220400461E}.}
\label{fig:trend}
\end{figure}
\end{center}

The mass asymmetry in the second chirp mass segment is in contrast to the mass ratio distribution for the binaries in the third chirp mass segment. All three models inferred comparable mass binaries in this segment. On comparing various draws of mass ratios predicted by the model $\mathbb{M}_{\mathcal{M}}$ for the second and the third segments, the draws from the third segment show greater mass symmetry 80\% of the time. On the other hand, the average mass ratio in the first segment is approximately the same as the second segment, however, the first segment contains highly asymmetric observations. The observations GW190814 and GW200210\_092254 have a chirp mass consistent with the first segment and have the lowest mass ratio among all the observations~\citep{o3b_cat}. Fig.~\ref{fig:trend} shows mass ratio becomes increasingly asymmetric with decreasing chirp mass. Similar to the case of the second peak, which is significant in the chirp mass but not in primary mass, the trend in the mass ratio shows dependence on the chirp mass but not on the primary mass. This is because the primary masses of the observations GW190814 and GW200210\_092254 are not consistent with any of the primary mass segments. 

Chirp mass dominates the phase evolution of \acp{GW} and thus is measured most accurately. Its measurement is also least affected by any systematic biases. However, chirp mass is not a parameter of astrophysical significance. The systematic dependence of mass ratio on chirp mass, but not primary or component mass is challenging to explain and will possibly result in significant implications. In addition, implications made using different mass parameters may be disparate. For example, the lack of observations in the chirp mass range 10--12$M_\odot$ and four well-placed peaks in the chirp mass distribution may be simply explained by a hierarchical merger scenario~\citep{2021ApJ...913L..19T, 2022ApJ...928..155T}. In this proposal, the first peak is created by the merger of \acp{BH} which are stellar remnants. The remnant \acp{BH} from previous \ac{BBH} merge further and create the next peaks. Of particular note is the relative location of peaks bearing a factor of 1.9, correctly accounting for the doubling of masses and around 5\% of it radiated in \ac{GW}\footnote{This proposal needs to explain why sub-dominant peak -- between dominant peaks -- are not visible in the inferred chirp mass distribution.}. But, the same implication is relatively difficult to make using the primary mass distribution as the confidence in the existence of the second peak is only marginal.

Although more observations are needed to improve confidence in the potential trend in the mass ratio, the trend is nevertheless intriguing. Future observations will only maintain the trend if there are \textbf{continued observations of binaries with chirp masses consistent with the first chirp mass segment but with large mass asymmetry}.

\section{Conclusion}
\label{sec:conclusion}

In this article, we reported our confidence in the four emerging peaks in the \ac{BH} mass distribution. The confidence in the first and the third peaks is significant, and confidence in the fourth peak is marginal. Interestingly, the confidence in the second peak depends on the mass parameter we choose to model the population on. When modelling the population on primary mass, the confidence in the second peak is marginal, but when modelling on the chirp mass the confidence is significant. Although chirp mass is not a parameter of astrophysical importance, the presence of a significant feature in the chirp mass distribution should not be ignored. We also reported a potential trend in the mass ratio distribution. Binaries are of comparable masses for the third peak, the second peak shows mass asymmetry and some of the observations consistent with the first peak have the largest mass asymmetry in the population.

\section*{Acknowledgements}

%We sincerely thank $\cdots$.

Sincere thanks to Stephen Fairhurst and Thomas Dent for helpful feedback on the manuscript. This work is supported by the STFC grant ST/V005618/1. We are grateful for the computational resources provided by Cardiff  University and funded by the STFC grant ST/N000064/1.

This research has made use of data, software and/or web tools obtained from the Gravitational Wave Open Science Center (https://www.gw-openscience.org/), a service of LIGO Laboratory, the LIGO Scientific Collaboration and the Virgo Collaboration. LIGO Laboratory and Advanced LIGO are funded by the United States National Science Foundation (NSF) as well as the Science and Technology Facilities Council (STFC) of the United Kingdom, the Max-Planck-Society (MPS), and the State of Niedersachsen/Germany for support of the construction of Advanced LIGO and construction and operation of the GEO600 detector. Additional support for Advanced LIGO was provided by the Australian Research Council. Virgo is funded, through the European Gravitational Observatory (EGO), by the French Centre National de Recherche Scientifique (CNRS), the Italian Istituto Nazionale della Fisica Nucleare (INFN) and the Dutch Nikhef, with contributions by institutions from Belgium, Germany, Greece, Hungary, Ireland, Japan, Monaco, Poland, Portugal, Spain.

%%%%%%%%%%%%%%%%%%%%%%%%%%%%%%%%%%%%%%%%%%%%%%%%%%
\section*{Data Availability}

The result files and plotting scripts are available on \href{https://github.com/vaibhavtewari/vamana/tree/653c47fa54576d878f59cf3ade44bbed1e56ff17}{GitHub}.

\section*{Packages used} 

\texttt{NumPy}~\cite{harris2020array}, \texttt{SciPy}~\cite{2020SciPy-NMeth}, \texttt{matplotlib}~\citep{Hunter:2007},  and \texttt{ASTROPY}~\citep{2022ApJ...935..167A}. 


\bibliographystyle{mnras}
\bibliography{references} % if your bibtex file is called example.bib

%\begin{thebibliography}{99}
%\end{thebibliography}

%%%%%%%%%%%%%%%%%%%%%%%%%%%%%%%%%%%%%%%%%%%%%%%%%%

%%%%%%%%%%%%%%%%% APPENDICES %%%%%%%%%%%%%%%%%%%%%
\appendix

\section{Method}
\label{method}

Inferring the compact binary population requires proper accounting of uncertainties in the estimated parameters, correcting bias caused due to the selective sensitivity of the detectors towards different binary parameters, and a model to infer the population~\citep{2019MNRAS.486.1086M, 2018ApJ...868..140T, 2019PASA...36...10T}. Given a model $p(\bm{\theta}|\Lambda)$ and a set of observations, $\bm{d} \equiv \{ \bm{d_0}, \cdots, \bm{d}_{N_\mathrm{obs}}\}$, the posterior on the model hyper-parameters is given by equation \ref{eq:pop_bayes},
\begin{equation}
p(\Lambda | \bm{d}) \propto
\prod_{i=1}^{N_\mathrm{obs}} \frac{\int d\bm{\theta}\;p(\bm{d}_{i} | \bm{\theta} )\;p (\bm{\theta} | \Lambda)}{\int \D\bm{\theta}\;p_{\mathrm{det}}(\bm{\theta})\; p(\bm{\theta}|\Lambda)} \;p(\Lambda),
\label{eq:pop_bayes}
\end{equation}
where $\bm{\theta}$ are the population parameters being inferred, $p(\Lambda)$ is the prior probability of the model hyper-parameters and $p_\mathrm{det}(\bm{\theta})$ encodes the probability of confidently observing a binary with parameters $\bm{\theta}$. The merger rate is estimated by extending Eq.~\ref{eq:pop_bayes} to include the Poisson probability of observing $N_\mathrm{obs}$ signals
\begin{equation}
    p(N_\mathrm{obs}|N_\mathrm{exp}) = N_\mathrm{exp}^{N_\mathrm{obs}}\;e^{-{N_\mathrm{exp}}},
\end{equation}
when the number of expected signals for a population model is $N_\mathrm{exp}$,
\begin{equation}
    N_\mathrm{exp} = \int \mathcal{R}(z)\;p(\bm{\theta}|\Lambda)\;\frac{\mathrm{d}V_c}{\mathrm{d}z} \mathrm{d}\bm{\theta}\;\mathrm{d}z.
\end{equation}
$\mathcal{R}(z)$, in the last equation, is the redshift-dependent merger rate and $\mathrm{d}V_c/\mathrm{d}z$ is the differential co-moving volume.

The posterior probability is Eq.~\ref{eq:pop_bayes} is estimated by \ac{MC} integration. The analysis estimating the parameters of \ac{GW} signals calculate $p(\bm{d}_{i} | \bm{\theta})$ using an independent Bayesian framework,  
\begin{equation}
    p (\bm{\theta} | \Lambda_{\mathrm{PE}}) \propto p(\bm{d}_{i} | \bm{\theta})\; p(\bm{\theta}|\Lambda_{\mathrm{PE}}).
\end{equation}
But, these analyses use a standard prior, $p(\bm{\theta}|\Lambda_{\mathrm{PE}})$, for all the signals. Furthermore, large-scale injection campaigns are performed to estimate the sensitivity of the detector network for a population model $p(\bm{\theta} | \Lambda_{\mathrm{inj}})$. Both the numerator and the denominator in Eq.~\ref{eq:pop_bayes} are then calculated for a target population $p (\bm{\theta} | \lambda)$ using importance sampling \citep{2019MNRAS.486.1086M, 2018CQGra..35n5009T}. Vamana uses the Metropolis-Hastings algorithm to sample the hyper-parameter posterior~\citep{10.1093/biomet/57.1.97}. Once hyper-parameter posteriors have been sampled, the inferred distribution on the mass parameter, $m$, can be obtained for each posterior by marginalising over the remaining parameters~$( \theta= (m, \omega))$,
\begin{equation}
p_i(m) = \int \mathrm{d}\omega\;p(m, \omega|\Lambda_i),
\label{eq:post_pred}
\end{equation}
where $i$ in $\Lambda_i$ identifies a sample.

Following, we list the three models presented in this article, Table~\ref{tab:Lambda} list the hyper-parameters for these models and their prior distributions:

i) Model $\mathbb{M}_{\mathcal{M}}$: Infers the chirp mass, mass ratio and aligned spin distributions. Each component consists of a truncated bi-variate normal to infer the chirp mass and mass ratio together and a uni-variate normal to identically but independently infer the aligned spin distribution. The mean chirp mass distribution inferred for a flat likelihood approximately follows a uniform-in-log distribution, i.e. $p(\langle\mathcal{M}|\Lambda\rangle) = \sum_ip(\mathcal{M}|\Lambda_i)/n \sim 1/\mathcal{M}$. The mean mass ratio inferred for a flat likelihood approximately follows, $p(\langle q|\Lambda\rangle) = \sum_i p(q|\Lambda_i)/n\sim q$. For this model's definition
\begin{multline}
p(\mathcal{M}, q, \chi_1, \chi_2, z|\Lambda) = \sum_{i=1}^N w_i \;\mathcal{N}(\mathcal{M}, q|\mu_i^{\mathcal{M}}, \sigma_i^{\mathcal{M}}, \mu_i^q, \sigma_i^q, C_i^{\mathcal{M}q})\\\phi(\chi_1|\mu^\chi_i, \sigma^\chi_i)\;\phi(\chi_2|\mu^\chi_i, \sigma^\chi_i),
\label{eq:model1}
\end{multline}
the bi-variate normal in Eq.~\ref{eq:model1} has mean and covariance matrix given as, 
\begin{equation}
    \mathrm{mean} = (\mu_i^{\mathcal{M}}, \mu_i^q), \; \mathrm{covariance} = \begin{bmatrix}
\left(\sigma_i^{\mathcal{M}}\right)^2 & C_i^{\mathcal{M}q}\\
C_i^{\mathcal{M}q} & \left(\sigma_i^q\right)^2
\end{bmatrix}
\end{equation}
with the distribution truncated outside values $0.05>q>1$.
\\\\
\noindent
ii) Model $\mathbb{M}_{m_1}$: Infers the primary mass, mass ratio and aligned spin distributions. Each component consists of a bi-variate normal to infer the primary mass and mass ratio together and a uni-variate normal to identically but independently infer the aligned spin distribution. The mean primary mass distribution inferred for a flat likelihood approximately follows a uniform-in-log distribution, i.e. $p(\langle m_1|\Lambda\rangle) = \sum_ip(m_1|\Lambda_i)/n \sim 1/m_1$. The mean mass ratio inferred for a flat likelihood approximately follows, $p(\langle q|\Lambda\rangle) = \sum_i p(q|\Lambda_i)/n\sim q$. For this model's definition,
\begin{multline}
p(m_1, q, \chi_1, \chi_2, z|\Lambda) = \sum_{i=1}^N w_i \;\mathcal{N}(m_1, q|\mu_i^{m_1}, \sigma_i^{m_1}, \mu_i^q, \sigma_i^q, C_i^{m_1q})\\\phi(\chi_1|\mu^\chi_i, \sigma^\chi_i)\;\phi(\chi_2|\mu^\chi_i, \sigma^\chi_i),
\label{eq:model2}
\end{multline}
the bi-variate normal in Eq.~\ref{eq:model2} has mean and covariance matrix given as, 
\begin{equation}
    \mathrm{mean} = (\mu_i^{m_1}, \mu_i^q), \; \mathrm{covariance} = \begin{bmatrix}
\left(\sigma_i^{m_1}\right)^2 & C_i^{m_1q}\\
C_i^{m_1q} & \left(\sigma_i^q\right)^2
\end{bmatrix},
\end{equation}
with the distribution truncated outside values $0.05>q>1$.
\\\\
iii) Model $\mathbb{M}^{pl}_{m_1}$: Infers the primary mass, mass ratio and aligned spin distributions. Each component consists of a uni-variate normal to infer the primary mass and a uni-variate normal to identically but independently infer the aligned spin distribution. Unlike models, $\mathbb{M}_{\mathcal{M}}$ and $\mathbb{M}_{m_1}$, model $\mathbb{M}^{pl}_{m_1}$ does not infer the mass ratio using a mixture model. Instead, it uses a single power law to model the mass ratio throughout the full mass range. The mean primary mass distribution inferred for a flat likelihood approximately follows a uniform-in-log distribution, i.e. $p(\langle m_1|\Lambda\rangle) = \sum_ip(m_1|\Lambda_i)/n \sim 1/m_1$. This model is defined as,
\begin{multline}
p(m_1, q, \chi_1, \chi_2, z|\Lambda) = \Big[\sum_{i=1}^N w_i \;\phi(m_1|\mu_i^{m_1}, \sigma_i^{m_1})\;\phi(\chi_1|\mu^\chi_i, \sigma^\chi_i)\\\phi(\chi_2|\mu^\chi_i, \sigma^\chi_i)\Big]\;\mathcal{P}(q|\alpha_i^q, q_i^{min}, 1.0).
\label{eq:model3}
\end{multline}
The older version of Vamana inferred the chirp mass and mass ratio distributions using a uni-variate normal and a power law respectively. Thus, the mass ratio did not vary with chirp mass within a component. Models, $\mathbb{M}_{\mathcal{M}}$ and $\mathbb{M}_{m_1}$ use truncated bi-variate normal distributions that have a correlation term included between the two parameters and thus the mass ratio can vary with the mass parameter within a component. All the models include a power law distribution in each component to infer the redshift evolution of the merger rate,
\begin{equation}
    \mathcal{R}(z) \propto (1 + z) ^{\kappa_i}
\end{equation}

The \ac{BBH} population has been inferred using the observations with a false alarm rate of at most once per year. We only used observations reported by the LVK collaborations~\citep{2019PhRvX...9c1040A, 2021PhRvX..11b1053A, 2021arXiv210801045T, o3b_cat}. Only observations with a mean chirp mass greater than 5$M_\odot$ are used. We excluded GW190814 from the results presented in Section~\ref{sec:results}. The total number of observations chosen is 69. For Fig.~\ref{fig:trend} we used a lower false alarm rate of at most twice per year and included all of the observations with mean chirp mass greater than 5$M_\odot$. This resulted in the inclusion of GW190814, GW190926, and GW200210\_092254. We used ten components for all three models. However, our results remain essentially unchanged for any number of components between 8 and 15. The binary parameters are estimated in the detector frame; to change to the source frame quantities we assume the Planck15 cosmology \citep{2016A&A...594A..13P}.

\def\arraystretch{1.3}
\begin{table*}
	\centering
	\caption{This table lists the hyper-parameters of the three models used to infer the \ac{BBH} population. The last column identifies the model where the hyper-parameters are used. U stands for Uniform, and UL for Uniform-in-log.}

	\begin{tabular}{lcccc} % four columns, alignment for each
		\hline
		$\Lambda$ & Description & Prior & Range & Models\\
		\hline
		$w_i$ & Mixing weights & Dirichlet($\bm{\alpha}$), $\alpha_{1\cdots N}= 1/N$  & (0, 1) & $\mathbb{M}_{\mathcal{M}}$, $\mathbb{M}_{m_1}$, $\mathbb{M}^{pl}_{m_1}$\\
		$\mu_i^{\mathcal{M}}$ & Location of Gaussians modeling chirp mass& UL & (5$M_\odot$,\; 60$M_\odot$) & $\mathbb{M}_{\mathcal{M}}$\\
		$\sigma_i^{\mathcal{M}}$ & Scale of Gaussians modeling chirp mass & U & (0.01\,$\mu_i^{\mathcal{M}}$,\; 0.15\,$\mu_i^{\mathcal{M}}$)\,/\,$\sqrt{N}$ & $\mathbb{M}_{\mathcal{M}}$\\
		$\mu_i^q$ & Location of Gaussians modeling mass ratio& UL & (.05, 1.04) & $\mathbb{M}_{\mathcal{M}}$\\
		$\sigma_i^q$ & Scale of Gaussians modeling mass ratio & U & (0.1,\; 1.0\,/\,$\sqrt{N}$) & $\mathbb{M}_{\mathcal{M}}$\\
  $C_i^{\mathcal{M}q}$ & Covariance between chirp mass and mass ratio& U & (-0.5\,$\sigma_i^{\mathcal{M}}\,\sigma_i^q$,\; 0.5\,$\sigma_i^{\mathcal{M}}\,\sigma_i^q$)& $\mathbb{M}_{\mathcal{M}}$\\
  $\mu^\chi_i$&Mean of the Gaussians modeling the aligned spin distribution&U for $|\chi_i| < 0.4$& UL for 0.4 < $|\chi_i|$ < 0.9& $\mathbb{M}_{\mathcal{M}}$, $\mathbb{M}_{m_1}$, $\mathbb{M}^{pl}_{m_1}$\\
  $\sigma^\chi_i$&Scale of the Gaussians modeling the aligned spin distribution&U&(0.1, 1.3\,/\,$\sqrt{N}$)& $\mathbb{M}_{\mathcal{M}}$, $\mathbb{M}_{m_1}$, $\mathbb{M}^{pl}_{m_1}$\\
  $\kappa_i$&Powerlaw exponent for the redshift evolution of the merger rate&U for $|1 + \kappa_i| < 0.1$& UL for 0.1 < $|1+\kappa_i|$ < 10 & $\mathbb{M}_{\mathcal{M}}$, $\mathbb{M}_{m_1}$, $\mathbb{M}^{pl}_{m_1}$\\
		\hline
		$\mu_i^{m_1}$ & Location of Gaussians modeling primary mass& UL & (5$M_\odot$,\; 80$M_\odot$) & $\mathbb{M}_{m_1}$\\
		$\sigma_i^{m_1}$ & Scale of Gaussians modeling primary mass & U & (0.01\,$\mu_i^{m_1}$,\; 0.2\,$\mu_i^{m_1}$)\,/\,$\sqrt{N}$ & $\mathbb{M}_{m_1}$\\
  $C_i^{m_1q}$ & Covariance between primary mass and mass ratio& U & (-0.5\,$\sigma_i^{m_1}\,\sigma_i^q$,\; 0.5\,$\sigma_i^{m_1}\,\sigma_i^q$)& $\mathbb{M}_{m_1}$\\  
  \hline
  $q_i^{min}$&Minimum value of the mass ratio (maximum is one)& U&(0.05, 0.25)&$\mathbb{M}^{pl}_{m_1}$\\
  $\alpha_i^q$&Slope of the power-law modeling mass ratio&U&(-7, 2)&$\mathbb{M}^{pl}_{m_1}$\\
  \hline
  \hline
	\end{tabular}
 \label{tab:Lambda}
\end{table*}
\def\arraystretch{1}

\section{Simulated Mass Distribution}
\label{sec:sims}
The tendency of a model to create local minima or maxima depends on the distribution of the data and the bandwidth of the modelling components.  Gaussian mixtures are supposed to inherently create local maxima and local minima in the inferred population. To generate a statistic on the creation of over or under-dense regions we perform 1,000 simulation runs for the featureless models. Instead of inferring a full-fledged simulated population, we only focus on the mass distribution. We simulate a mass distribution that best describes the inferred mass distribution but contains no local maxima. To achieve that we
\begin{enumerate}
    \item fit a monotonically decreasing broken power-law, such that, the inferred mass distribution is least over-dense compared to it throughout the mass range. Table~\ref{tab:bpl} lists the best-fit parameters for the three models.
    \item apply selection effects and draw 69 data points from the resulting distribution\footnote{This involves applying importance sampling to $p(\bm{\theta} | \Lambda_{\mathrm{inj}})$ and obtaining samples from the target mass distribution. We just perform sampling on mass distribution, thus the mass-ratio and spin distributions of the synthetic population are the same as that of the injections~\citep{ligo-O1O2O3-search-sensitivity}.}.
    \item include measurement uncertainty by augmenting each data point with additional 500 samples drawn from a normal distribution. The values of the data points serve as the mean for these normal distributions and the scales are empirically estimated depending on the mass value. We found the following relation best describes the variation of measurement uncertainty with the mass value\footnote{This is only an approximation. In reality, the measurement uncertainty depends on other factors, such as signal-to-noise ratio.},
    \begin{equation}
        \frac{\sigma(\mathcal{M})}{\mathcal{M}} = 0.023 + 0.002\,\mathcal{M},\quad
        \frac{\sigma(m_1)}{m_1} = 0.167 - 0.00047\,m_1.
    \end{equation}
    \item infer the mass distribution using the simulated data. Priors identical to models $\mathbb{M}_{\mathcal{M}}$, $\mathbb{M}_{m_1}$, and $\mathbb{M}^{pl}_{m_1}$ are used in inferring the chirp or primary mass distribution for the featurelss models.
    \item estimate $f$ for each simulation. For each simulation record the highest value, $f_{\mathrm{highest}}$, obtained within each mass segment. This results in one thousand $f_{\mathrm{highest}}$ values for each mass segment.
\end{enumerate}

\def\arraystretch{.5}
\begin{table}
	\centering
	\caption{The best-fit parameters of the monotonically decreasing broken power-law, $f(x) \propto x^{-\alpha_i} \;\text{for}\; x_{i}\leq x < x_{i+1}$. The population model's inferred mass distribution is least dense compared to this distribution throughout the mass range. The table lists the endpoints~($x_i$) and exponents~($\alpha_i$) of the broken power-law. All endpoints have units of solar mass.}
	\begin{tabular}{ccccccccccc} % four columns, alignment for each
  \hline
      &$x_i$ & 4 & & 10 & & 18 & & 37 & & 65\\
      $\mathbb{M}_{\mathcal{M}}$\\
      &$\alpha_i$ & & 4.66 & & 0.67 & & 2.17 & & 4.58 &\\
  \hline
      &$x_i$ & 5 & & 14 & & 25 & & 50 & & 90\\
      $\mathbb{M}_{m_1}$\\
      &$\alpha_i$ & & 3.86 & & 0.01 & & 3.70 & & 3.98 &\\
  \hline
      &$x_i$ & 5 & & 14 & & 25 & & 50 & & 90\\
      $\mathbb{M}^{pl}_{m_1}$\\
      &$\alpha_i$ & & 3.93 & & 0.54 & & 3.09 & & 4.93 &\\
  \hline
	\end{tabular}
 \label{tab:bpl}
\end{table}

We ignored selection effects in the simulations. Thus, our underlying population for the calculation of $f$ is the selection-weighted broken power-law. We expect the inclusion of selection effects will only impact the weights of the components. The components modelling the lower masses will have their weights increased compared to analysis that ignores the selection effect. We indirectly verified this by inferring the \emph{observed} \ac{BBH} distribution~(i.e. ignoring correction of the selection effects thus just modelling the observed mass and spin distribution of \acp{BBH} in Earth's frame) using the three models. Our conclusion remains unchanged on the significance of the peaks as the value of $f_{\mathrm{highest}}$ reported in Table~\ref{tab:peaks_summary} remained mostly unchanged. 

%%%%%%%%%%%%%%%%%%%%%%%%%%%%%%%%%%%%%%%%%%%%%%%%%%

% Don't change these lines
\bsp	% typesetting comment
\label{lastpage}
\end{document}